\documentclass[11pt,floatfix]{revtex4}
\usepackage[dvips]{graphicx,color}
\usepackage{amsmath}
\usepackage{amsfonts}
\usepackage[latin1]{inputenc}

\def\sint{\int \!\!\!\!\!\!\!\!\!\sum_{\ \ np}\ }

\begin{document}

\title{Generalized Ginzburg-Landau approach to inhomogeneous phases in nonlocal chiral quark models}

\author{J.P.\ Carlomagno$^{a,b}$, D.\ G\'omez Dumm$^{a,b}$ and N.N.\ Scoccola$^{b,c,d}$}

\address{$^{a}$ IFLP, CONICET $-$ Dpto.\ de F\'{\i}sica, FCE, Universidad
Nacional de La Plata, C.C. 67, 1900 La Plata, Argentina,\\
$^{b}$ CONICET, Rivadavia 1917, 1033 Buenos Aires, Argentina \\
$^{c}$ Physics Department, Comisi\'on Nacional de Energ\'{\i}a Atómica,
Av.Libertador 8250, 1429 Buenos Aires, Argentina \\
$^{d}$ Universidad Favaloro, Sol{\'\i}s 453, 1078 Buenos Aires, Argentina}

\begin{abstract}
We analyze the presence of inhomogeneous phases in the QCD phase
diagram within the framework of nonlocal chiral quark models. We
concentrate in particular in the positions of the tricritical
(TCP) and Lifshitz (LP) points, which are studied in a general
context using a generalized Ginzburg-Landau approach. We
find that for all the phenomenologically acceptable model
parametrizations considered the TCP is located at a higher
temperature and a lower chemical potential in comparison with the
LP. Consequently, these models seem to favor a scenario in which
the onset of the first order transition between homogeneous phases
is not covered by an inhomogeneous, energetically favored phase.
\end{abstract}

\maketitle

\hfill

The behavior of strongly interacting matter under extreme
conditions of temperature and/or density has been extensively
studied along the last decades. However, after a considerable
amount of theoretical and experimental work, the phase diagram of
Quantum Chromodynamics (QCD) still remains poorly understood. For
instance, qualitative features such as the precise nature of the
chiral phase transition at low temperatures, or even the existence
of a critical point, have not been firmly established yet. From
the theoretical point of view, one of the main reasons for this
state of affairs is that the ab-initio lattice QCD approach has
difficulties to deal with the region of medium/low temperatures
and moderately high densities, owing to the so-called ``sign
problem''. Thus, most of the present knowledge about the behavior
of strongly interacting matter arises from the study of effective
models, which offer the possibility to get predictions of the
transition features at regions that are not accessible through
lattice techniques. In this context, in the last years some works
have considered that the chiral symmetry restoration at low
temperatures could be driven by the formation of non-uniform
phases~\cite{Buballa:2014tba}. One particularly interesting result
suggests that the expected critical endpoint of the first order
chiral phase transition might be replaced by a so-called Lifshitz
point (LP), where two homogeneous phases and one inhomogeneous
phase meet~\cite{Nickel:2009ke}. This result has been obtained in
the chiral limit ---where the end point becomes a tricritical
point (TCP)--- in the framework of the well-known
Nambu$-$Jona-Lasinio model (NJL)~\cite{njl}, in which quark fields
interact through a local chiral invariant four-fermion coupling.
More recently, this issue has also been addressed in the context
of a quark-meson model with vacuum
fluctuations~\cite{Carignano:2014jla}, where it is found that the
LP might coincide or not with the TCP depending on the model
parametrization.

The aim of the present work is to analyze the relation between the positions
of the TCP and the LP in the framework of nonlocal chiral quark models.
These theories are a sort of nonlocal extensions of the NJL model, and
intend to represent a step towards a more realistic modelling of QCD. In
fact, nonlocality arises naturally in the context of successful approaches
to low-energy quark dynamics~\cite{Schafer:1996wv,RW94}, and it has been
shown~\cite{Noguera:2008} that nonlocal models can lead to a momentum
dependence in the quark propagator that is consistent with lattice QCD
results~\cite{bowman,Parappilly:2005ei,Furui:2006ks}. Another advantage of
these models is that the effective interaction is finite to all orders in
the loop expansion, and therefore there is no need to introduce extra
cutoffs~\cite{Rip97}. Moreover, in this framework it is possible to obtain
an adequate description of the properties of strongly interacting particles
at both zero and finite
temperature/density~\cite{Noguera:2008,Bowler:1994ir,Schmidt:1994di,Golli:1998rf,
GomezDumm:2001fz,Scarpettini:2003fj,GomezDumm:2004sr,GomezDumm:2006vz,
Hell:2008cc,Contrera:2009hk,Hell:2009by,Contrera:2010kz,Dumm:2010hh,
Pagura:2011rt,Carlomagno:2013ona}.

We consider here the simplest version of a nonlocal SU(2) chiral
quark model in the chiral limit. The corresponding Euclidean
effective action is given by
\begin{equation}
S_{E}= \int d^{4}x\
\left[ -i \bar{\psi}(x) \; \rlap/\partial \; \psi(x)
-\frac{G}{2} \ j_{a}(x) \; j_{a}(x)  \right] \ ,
\label{action}%
\end{equation}
where $\psi$ stands for the $N_{f}=2$ fermion doublet
$\psi\equiv(u,d)^T$. The nonlocal currents $j_{a}(x)$ are given by
\begin{align}
j_{a}(x)  &  =\int d^{4}z\ {\cal G}(z)\ \bar{\psi}\left(x+\frac{z}{2}\right)
\ \Gamma_{a}\ \psi\left(  x-\frac{z}{2}\right) , \label{currents}%
\end{align}
where we have defined
$\Gamma_{a}=(\leavevmode\hbox{\small1\kern-3.8pt\normalsize1},i\gamma
_{5}\vec{\tau})$, and the function ${\cal G}(z)$ is a nonlocal
form factor that characterizes the effective interaction.

To proceed we perform a standard bosonization of the theory, in
which bosonic fields are introduced and quark fields are
integrated out. We will work within the mean field approximation,
replacing the bosonic scalar and pseudoscalar fields by their
vacuum expectation values $\sigma (\vec x)$ and $\pi_a (\vec x)$,
respectively. The mean field values are allowed to be
inhomogeneous, hence the explicit dependence on spatial
coordinates. The resulting mean field Euclidean action reads then
\begin{eqnarray}
S_{E} = - \,\mbox{Tr}\; \log \ S^{-1} \; + \; \frac{1}{2G} \int d^3 x \;
\phi^a(\vec x)\, \phi^a(\vec x)\ ,
\end{eqnarray}
where we have introduced the chiral four-vector $\phi^a = (\sigma (\vec x) ,
\vec \pi(\vec x))$, and the operator $S^{-1}$ is given by
\begin{eqnarray}
S^{-1}(x,y) \; = \;\delta^{4}(x-y) \; (-i \rlap/\partial_y) \; +
\;{\cal G}(x-y) \; \Gamma^a \, \phi^a \left((\vec x + \vec
y)/2\right) \ .
\end{eqnarray}

The extension to finite temperature $T$ and chemical potential $\mu$ can be
performed by following the usual Matsubara procedure. Once the operators are
transformed to momentum space, for a given integral of any operator $F$ over
the fourth component of the momentum ($p_4$) we carry out the replacement
\begin{eqnarray}
\int \frac{dp_4}{2 \pi}\; F[p_4,\dots]\ \ \rightarrow \ \ T
\sum_{n=-\infty}^{\infty} F[(2 n+1) \pi T - i \mu, \dots] \ ,
\end{eqnarray}
where the dots stand for other variables on which $F$ might depend upon.

As stated, we are interested in the determination of the Lifshitz point (LP)
---i.e., the point where the inhomogeneous phase and the two homogeneous
phases with broken and restored chiral symmetry meet--- and its
location relative to the tricritical point (TCP) in the $(T,\mu)$
plane. If the analysis is restricted to homogeneous phases, in the
chiral limit the TCP denotes the point where the second-order
chiral phase transition turns into a first order one. We will
consider here the so-called Ginzburg-Landau (GL) approach, in
which the relative positions of the LP and TCP can be analyzed in
a rather general way that does not require to specify the explicit
form of the inhomogeneity~\cite{Nickel:2009ke,Abuki:2011pf}. We
follow the analysis proposed in Ref.~\cite{Nickel:2009ke}, where
the mean field thermodynamic potential is expanded around the
symmetric ground state in powers of the order parameters and their
spatial gradients. Let us carry this double expansion up to sixth
order, i.e.~up to terms with coefficients carrying dimensions
(energy)$^{-2}$. The GL functional should have the general
form~\cite{Iwata:2012bs}
\begin{eqnarray}
\omega(T,\mu,\phi^a(\vec x)) & = & \frac{\alpha_2}{2} \ \phi^2
+ \frac{\alpha_4}{4} (\phi^2)^2 + \frac{\alpha_{4b}}{4} (\nabla \phi)^2 +
\nonumber \\
& &
\frac{\alpha_6}{6} (\phi^2)^3 + \frac{\alpha_{6b}}{6} (\phi, \nabla \phi)^2
+ \frac{\alpha_{6c}}{6} \left[ \phi^2 (\nabla \phi)^2 - (\phi, \nabla \phi)^2 \right]
+ \frac{\alpha_{6d}}{6} \ (\triangle \phi)^2 \ ,
\label{genome}
\end{eqnarray}
where $\phi^2 = (\phi,\phi) = \phi_a \phi_a = \sigma^2 + \vec \pi \ \! ^2$,
$(\phi, \nabla \phi) = \phi_a \nabla\phi_a = \sigma \nabla \sigma + \vec \pi
\nabla \vec \pi$, etc.

In the particular case of the nonlocal models considered in this work, a
somewhat lengthy but straightforward calculation leads to the following form
for the GL coefficients:
\begin{eqnarray}
\alpha_2 &=& \frac{1}{G} - 8\; N_c\ \sint \frac{g^2}{p_n^2}
\nonumber \\
\alpha_4 &=& 8\; N_c\ \sint \frac{g^4}{p_n^4}
\nonumber \\
\alpha_{4b} &=& 8 \; N_c\ \sint \frac{g^2}{p_n^4} \left(1 -
\frac{2}{3}\,\frac{g'}{g}\, \vec p^{\;2}\right) \nonumber
\end{eqnarray}
\begin{eqnarray}
\alpha_6 &=& - 8 \; N_c\ \sint \frac{g^6}{p_n^6}
\nonumber \\
\alpha_{6b} &=& -40 \; N_c\ \sint \left[ \frac{g^4}{p_n^6} \left( 1 -
\frac{26}{15}\,\frac{g'}{g}\, \vec p^{\; 2}
 + \frac{8}{5} \, \frac{{g'}^2}{g^2}\, \vec p^{\; 2} p_n^2 \right) \right]
 \nonumber \\
\alpha_{6c} &=& -24 \; N_c\ \sint \frac{g^4}{p_n^6} \left( 1 - \frac{2}{3}\,\frac{g'}{g}\,
\vec p^{\; 2} \right)
\nonumber \\
\alpha_{6d} &=& -4 \; N_c\ \sint \frac{g^2}{p_n^6}
\left[ 1 - \frac{2}{3}\,\frac{g'}{g}\,\vec p^{\; 2} + \frac{1}{5}
\left(\frac{{g'}^2}{g^2}+\frac{g''}{g}
\right)\vec p^{\; 4}\right] \ ,
\label{results}
\end{eqnarray}
where we have used the shorthand notation
\begin{equation}
\sint  \equiv \ \frac{T}{2 \pi^2}\sum_{n=-\infty}^{\infty}
\int_0^\infty d|\vec p|\; \vec p^{\; 2} \ , \label{nonloc}
\end{equation}
and $p_n^2 \equiv \left[ (2n+1) \pi T - i \, \mu \right]^2 + \vec
p^{\; 2}$. The function $g$ is the Fourier transform of the form
factor ${\cal G}(x)$ (which for the moment is only assumed to be
invariant under spatial rotations) evaluated at $p^2 = p_n^2$,
while $g'$, $g''$ denote derivatives with respect to $\vec p^{\;
2}$. It should be noted that, except for those in $\alpha_{6b}$,
all the derivatives appearing in these expressions can be
eliminated through integration by parts. We have chosen to present
the results in the above given form so as to facilitate the
comparison with the NJL results quoted in
Ref.~\cite{Nickel:2009ke}, which should correspond to ${\cal G}(x)
= \delta^{(4)}(x)$, i.e.~$g=1$. Indeed, in this limit, from
Eqs.~(\ref{results}) one gets
\begin{eqnarray}
& \alpha_{4b} = \alpha_4 \ , &
\label{njla4} \\
& \alpha_{6b}/5 = \alpha_{6c}/3 = 2\, \alpha_{6d} = \alpha_6 \ ,&
\label{njla6}
\end{eqnarray}
in agreement with Refs.~\cite{Nickel:2009ke,Iwata:2012bs}. A
regularization prescription has to be also introduced in order to
avoid ultraviolet divergences.

We turn now to the main topic of this work, namely, the predictions of
chiral quark models for the relative positions of the tricritical and
Lifshitz points in the $(T,\mu)$ plane. By looking at the GL functional in
Eq.~(\ref{genome}), it is seen that for $\alpha_{4b} > 0$ the system is in
the usual homogeneous phase. Now if in addition one has $\alpha_4 > 0$, the
system undergoes a first order chiral restoration transition when
$\alpha_2=0$ ($\phi^2=0$ for $\alpha_2>0$, $\phi^2\neq 0$ for $\alpha_2<0$),
which defines a first order transition line in the $T-\mu$ plane. This line
ends at the tricritical point, where also $\alpha_4=0$ is satisfied. Thus
the position of the TCP can be determined by solving the set of equations
\begin{equation}
\alpha_2 = 0 \ , \qquad \alpha_4 = 0 \ .
\label{eqtcp}
\end{equation}
On the other hand, for $\alpha_{4b} < 0$ inhomogeneous solutions are
favored. Hence the Lifshitz point, i.e., the point where the onset of the
inhomogeneous phase meets the chiral transition line, is obtained
from~\cite{Buballa:2014tba}
\begin{equation}
\alpha_2 = 0 \ , \qquad \alpha_{4b} = 0 \ .
\label{eqlp}
\end{equation}
It is clear from Eq.~(\ref{njla4}) that within the NJL model the TCP and the
LP are predicted to coincide. However, given the differences between the
expressions for $\alpha_4$ and $\alpha_{4b}$ in Eqs.~(\ref{results}), there
is no reason to expect this coincidence to hold in the framework of nonlocal
models. In order to determine the relative position of the TCP and LP within
these models we have to solve Eqs.~(\ref{eqtcp}) and (\ref{eqlp}). This can
be done numerically once we have taken some model parametrization, i.e., a
set of values for the model parameters and a definite shape for the form
factor. We start by choosing the Gaussian form
\begin{equation}
g = \exp(- p^2/\Lambda^2) \ ,
\label{gaussian}
\end{equation}
which has been frequently considered in the
literature~\cite{Bowler:1994ir,Schmidt:1994di,Golli:1998rf,
GomezDumm:2001fz,Scarpettini:2003fj,GomezDumm:2004sr,GomezDumm:2006vz}.
Notice that the form factor introduces a parameter $\Lambda$ that
indicates the range of the interaction in momentum space. Thus, in
the chiral limit, the model is completely determined by $\Lambda$
and the coupling constant $G$. It is usual to fix these parameters
so as to get phenomenologically adequate values for the pion decay
constant and the quark-antiquark condensate. Here, according to
the recent analysis in Ref.~\cite{Aoki:2013ldr}, we will take
$f_\pi^{\rm ch}=86$~MeV and $\langle\bar q q\rangle^{\rm ch} =
-(270$~MeV$)^3$ (superindices stress that values correspond to the
chiral limit). From dimensional analysis it is immediate to see
that any dimensionless quantity turns out to be just a function of
the dimensionless combination $\bar G = G \Lambda^2$, while
dimensionful quantities (such as e.g.\ the coordinates of the TCP
and LP in the $T-\mu$ plane) can be written as a function of $\bar
G$ times some power of a dimensionful parameter, say e.g.\ the
pion decay constant $f_\pi^{\rm ch}$. The ``physical'' value of
$\bar G$ will be that leading to a ratio $-(\langle\bar q
q\rangle^{\rm ch} )^{1/3}/f_\pi^{\rm ch} \simeq 3.14$, which
arises from the phenomenological values quoted above. Numerically
we obtain $G= 14.65$~GeV$^{-2}$, $\Lambda = 1.045$~GeV, $\bar G =
16.03$. In order to check the parameter dependence of our results
we will consider values for $-(\langle\bar q q\rangle^{\rm ch}
)^{1/3}/f_\pi^{\rm ch}$ in the range $3.0$ to $3.3$. For
$f_\pi^{\rm ch}=86$~MeV, this corresponds to a shift $\lesssim
10$~MeV around the central value $-(\langle\bar q q\rangle^{\rm
ch})^{1/3} = 270$~MeV.

Our numerical results for the coordinates of the TCP and LP are
displayed in Fig.~1. In the left panel we show the positions of
these points in the $T-\mu$ plane, for the mentioned range of
values of $-(\langle\bar q q\rangle^{\rm ch} )^{1/3}/f_\pi^{\rm
ch}$. Notice that values of $T$ and $\mu$ are normalized to units
of $f_\pi^{\rm ch}$. It is seen that for the considered parameter
range the LP is always found at a lower temperature and a larger
chemical potential than the TCP. In the right panel of Fig.~1 we
plot the ratio $-(\langle\bar q q\rangle^{\rm ch}
)^{1/3}/f_\pi^{\rm ch}$ as a function of the dimensionless
parameter $\bar G$. Here the dashed line indicates the
``physical'' value mentioned above.

\begin{figure}[hbt]
\includegraphics[width=0.7 \textwidth]{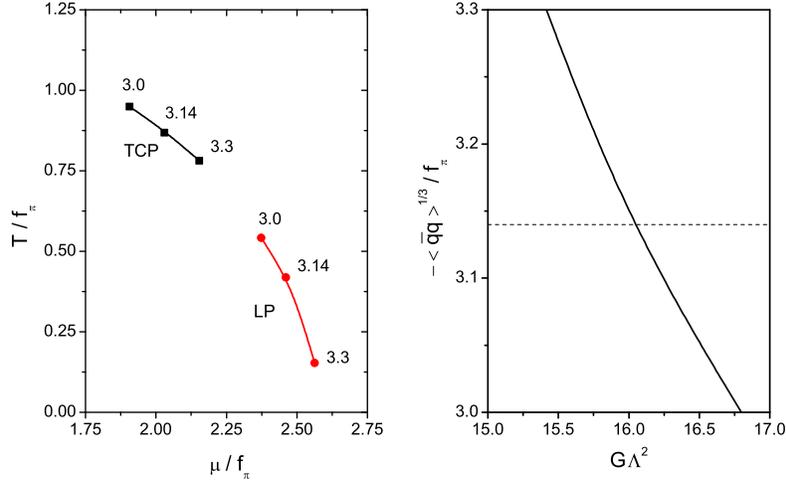}
\caption{Left: temperature and chemical potential for the TCP and
LP, in units of $f_\pi$, for different values of the ratio
$-(\langle\bar q q\rangle^{\rm ch} )^{1/3}/f_\pi^{\rm ch}$. Right:
relation between this ratio and the dimensionless parameter $\bar
G = G\,\Lambda^2$. The dashed line corresponds to the
phenomenologically preferred value $-(\langle\bar q q\rangle^{\rm
ch} )^{1/3}/f_\pi^{\rm ch}\simeq 3.14$.} \label{fig1}
\end{figure}

A somewhat better understanding of the results can be achieved by
taking into account approximate analytical expressions for the GL
coefficients. In fact, through the methods discussed in Appendix A
of Ref.~\cite{GomezDumm:2004sr} we obtain the relations
\begin{eqnarray}
\alpha_2 &=& \frac{1}{G} + \frac{N_c}{\pi^2} \left[
\frac{\pi^2}{3}\,T^2+\mu^2 - \int_0^\infty dp \ p\; g^2(p^2) \right]
\nonumber \\
\alpha_4 &=& - \frac{N_c}{\pi^2} \left[
4 g'(0)\left(\frac{\pi^2}{3}\,T^2+\mu^2\right) + \frac{1}{2} +
\log{2} - \int_0^\infty dp \ \frac{g^4(p^2) - n_+(p) - n_-(p)}{p} \right]
\nonumber \\
\alpha_{4b} &=& - \frac{N_c}{\pi^2} \left[\
g'(0)\left(\frac{\pi^2}{3}\,T^2+\mu^2\right) - \frac{2}{3}
\left[g'(0)^2+g''(0)\right]\;\left(\frac{7\pi^4}{30}\,T^4 +
\pi^2\,T^2\,\mu^2+\frac{1}{2}\mu^4\right) + \right. \nonumber \\
& & \qquad \qquad \left. \frac{3}{8} + \log{2} - \int_0^\infty dp
\ \frac{g^2(p^2) - n_+(p) - n_-(p)}{p}\ \right]\ .
\end{eqnarray}
The above expression for $\alpha_{4b}$ has not, to our knowledge,
been reported before, while those for $\alpha_2$ and $\alpha_4$
have been already given (using a different notation) in
Ref.~\cite{GomezDumm:2004sr}. One can check that in the region of
interest these relations provide a very good approximation (in
general, below the percent level) to the results arising from the
numerical evaluation of the Matsubara sums. In order to determine
the relative positions between the TCP and the LP, it is
interesting to calculate the coefficient $\alpha_{4b}$ at the TCP,
i.e.\ where $\alpha_2 = \alpha_4 = 0$. We obtain
\begin{eqnarray}
{\alpha_{4b}}^{\rm \tiny{(TCP)}} &=& \frac{N_c}{\pi^2} \left\{
\pi^2\,g'(0)\, T_c^2 + \frac{\pi^4}{9}\left[g'(0)^2+g''(0)\right]
\left[\frac{7}{5}\,T^4 + 2\,T^2(T_c^2-T^2)+\frac{1}{3}\,(T_c^2-T^2)^2
\right] + \right. \nonumber \\
& & \left.  \frac{1}{8} - 4 \int_0^\infty dp \ p\,\log p\; [1 -
2g^2(p^2)]\;g(p^2)\, g'(p^2) \ \right\} \ ,
\label{a4tcp}
\end{eqnarray}
where $T_c$ stands for the (second order) chiral phase transition
temperature at $\mu = 0$. From the GL expansion it is easy to see
that the condition $\alpha_{4b}^{\rm \tiny{(TCP)}} > 0$ $(< 0)$
implies that the LP is located at lower (higher) temperature and
higher (lower) chemical potential than those of the TCP. In the
particular case of the Gaussian form factor Eq.~(\ref{a4tcp})
reduces to
\begin{eqnarray}
\alpha_{4b}^{\tiny\rm (TCP)} &=& N_c \left\{ - \, t_c^2
 + \frac{1 + 4\,\log 2}{8\pi^2} + \frac{\pi^2}{9}
\left[\frac{14}{5}\,t^4 + 4\,t^2(t_c^2-t^2)+\frac{2}{3}\,(t_c^2-t^2)^2
\right]\right\} \ ,
\end{eqnarray}
where we have defined $t = T/\Lambda$, $t_c = T_c/\Lambda$. It can
be seen that in this case one can get $\alpha_{4b}^{\rm
\tiny{(TCP)}} < 0$ only if the dimensionless constant $\bar G$
satisfies
\begin{equation}
\bar G \; > \; \frac{4\pi^2}{N_c [\sqrt{6(13 - 2\log 2)} - 8]} \;\simeq\;
37.9\ ,
\end{equation}
which is far from the phenomenologically accepted range (see lower
right panel in Fig.~1).

For definiteness we have discussed so far the particular case of
the Gaussian nonlocal form factor in Eq.~(\ref{gaussian}). In
order to get an insight of whether the results can be extended to
other form factor shapes we have also considered the Lorentzian
functions
\begin{equation}
g = \frac{1}{1 + (p^2/\Lambda^2)^n} \ ,
\end{equation}
with $n \geq 2$. For $n=2$, which corresponds to a rather ``soft''
ultraviolet behavior, the situation concerning the relative
positions of the TCP and LP is found to be quite similar to that
of the Gaussian form factor. If $n$ is increased, both the TCP and
LP tend to be located at lower temperatures, and eventually the LP
disappears. In all phenomenologically acceptable cases the TCP is
found to be located at a higher temperature and a lower chemical
potential than those of the LP. It is also worth mentioning that
Eqs.~(\ref{results}) are also valid for ``instantaneous'' form
factors, i.e.~those that only depend on space variables, ${\cal
G}(|\vec x|)$. In general these form factors lead to rather large
values of the chiral condensate~\cite{Grigorian:2006qe}. Numerical
solutions of Eqs.~(\ref{eqtcp}) and (\ref{eqlp}) allow to find the
corresponding locations of the TCP and LP, which are qualitatively
similar to those obtained for the covariant form factors.

In conclusion, we have analyzed the relation between the positions
of the tricritical point (TCP) and the Landau point (LP) in the
framework of the simplest version of nonlocal chiral quark models
using the generalized Ginzburg-Landau approach. We have found that
for all the phenomenologically acceptable parametrizations
considered the TCP is located at a higher temperature and a lower
chemical potential in comparison with the LP. Consequently, these
models seem to favor a scenario in which the onset of the first
order transition between homogeneous phases is not covered by an
inhomogeneous, energetically favored phase. This differs from what
happens in the local NJL model, where the TCP and LP are predicted
to coincide~\cite{Nickel:2009ke}, or in quark-meson models with
vacuum fluctuations, where the relative position of these points
depends on the model parametrization~\cite{Carignano:2014jla}. The
location of the TCP and LP has also been investigated numerically
in a recent study based on the Dyson-Schwinger
approach~\cite{Muller:2013tya}. Although the corresponding result
seems to agree with that of the local NJL model, we should keep in
mind that a precise numerical determination of the positions of
the TCP and LP is in general a quite difficult task.

Several extensions of our work deserve further investigations. For
example, it would be important to incorporate isoscalar vector
meson interactions, to consider the coupling to the Polyakov loop
and to analyze the effect of wave function renormalization.
Moreover, the actual determination of the size of inhomogeneous
phases in the $(T,\mu)$ plane in the context of nonlocal models
should be feasible, at least for simple inhomogeneous
configurations. We expect to report on these issues in forthcoming
publications.

This work has been partially funded by CONICET (Argentina)
under grants PIP 00682 and PIP 00449, and
by ANPCyT (Argentina) under grant PICT11-03-00113.

\end{document}